\documentclass[useAMS,usenatbib,referee]{mn2e}

\usepackage{epsfig}

\title[Analytical framework for haloes and galaxies orientation]
{An analytical framework to describe the
orientation of dark matter halos and galaxies within the large-scale structure}

\author[Juan E. Betancort-Rijo and Ignacio Trujillo]{Juan E. Betancort-Rijo $^{1,2}$\thanks{E-mail:
jbetanco@iac.es; trujillo@iac.es} and Ignacio Trujillo$^{1,2}$\footnotemark[1]\\ $^{1}$Instituto de
Astrof\'isica de Canarias, E-38205, La Laguna, Tenerife, Spain\\ $^{2}$Departamento de Astrof\'isica, 
Universidad de La Laguna, E-38205 La Laguna, Tenerife, Spain}

\begin{document}

\date{Accepted 1988 December 15. Received 1988 December 14; in original form 1988 October 11}

\pagerange{\pageref{firstpage}--\pageref{lastpage}} \pubyear{2002}

\maketitle

\label{firstpage}

\begin{abstract}

We provide a set of general tools for studying the alignments of dark matter halos and
galaxies with respect to the large scale structure. The statistics of the positioning of
these objects is represented by a Probability Distribution Function (PDF) of their Euler
angles. The mathematical identities relating this PDF to the alignment of the halo axes with
respect to the direction that characterises locally the large scale structure are given. The
PDF corresponding to halos located in the shells of the cosmic voids is inferred from
previous results. This PDF is used to show how to recover the outcomes found for the
alignments of the axes of these halos in simulations. We also explore the orientation of the
angular momentum of the halos, both with respect to the halo axes and with respect to the
large scale structure. We present an expression which describes well numerical results for
the alignment of the angular momentum of the halo with respect to the halo axes for randomly
chosen  halos. Combining that expression, with the PDF of the halo positioning with respect
to the large scale structure we find, in the case of halos in the shells of voids, an
alignment of the angular momentum that is opposite to that found in simulations. To solve
this issue, we propose a model that relates the orientation of the angular momentum with the
halos axes accounting for the orientation of the halo axes with the large scale structure.
This model is shown to recover  accurately the observed PDF of the halo angular momentum with
respect to the void radial direction. This model also predicts a substantial dependence of
the intrinsic alignment of the angular momentum on environment.  In addition, we give an
expression for determining the degradation of the angular momentum intrinsic alignment when
observational errors are accounted. This expression is also used to determine the departure
of the observed value of the alignment from the initial expectation (as provided by the tidal
torque theory) due to the rotation of the angular momentum of the halo with respect to the
initial torque. For voids, we find that the strength of the alignment is reduced to half
the original value. We discuss how to adapt the void results to other cosmic large scale
structures (i.e. filaments, walls, etc).

\end{abstract}


\section{Introduction}

Our current galaxy formation paradigm states that galaxies acquire their angular momentum  by
a mechanism known as tidal torquing (Sciama 1955; Peebles 1969;  Doroshkevich 1970; White
1984). In this paradigm, tidal fields generated from the large-scale structure excerpt a
torquing on the protogalactic object prior to gravitational collapse. The acquired angular
momentum is conserved during collapse and their value will dictate the formation and
orientation of the galactic disks. One important consequence of the tidal torquing theory
(TTT) is an expected relation between the large-scale structure and the orientation of the
angular momentum of the galaxies within it (see a recent review by Sch\"afer 2009).

From the observational point of view, Navarro et al. (2004) found a notable (with a
significance of 92\%) excess of edge-on spiral galaxies (for which the orientation of the
disk rotation axis may be determined unambiguously) highly inclined relative to the
supergalactic plane. Trujillo et al. (2006) perform a measurement of the rotation axes of
galaxies situated on the shells of large cosmic voids and found that the angular momentum of
these objects tend to be aligned perpendicular to the radial direction of the void with a
significance of 99.7\% (see also Slosar \& White 2009). Lee \& Erdogdu (2007) combined a
reconstruction of the density tidal shear field from the 2MASS redshift survey
with the spin field derived from the Tully galaxy catalogue and found a correlation strength
at the 6$\sigma$ significance level.

From the numerical perspective, the orientation of the halos within their large-scale cosmic
structure has been investigated by several authors (Bailin \& Steinmetz 2005; Patiri et al.
2006; Brunino et al. 2007; Arag\'on-Calvo et al. 2007; Cuesta et al. 2008). In particular,
Patiri et al. (2006) and Brunino et al. (2007) investigated the alignment of dark matter
halos within the shells of large cosmic voids  and provided evidence that the minor axis of
the dark matter halos points preferentially to the centres of the voids. The intermediate
and major axis were found to be preferentially oriented perpendicularly to the void
direction. In addition, Cuesta et al. (2008) found, with high significance, a preferential
alignment between the angular momentum of the halo and the void direction. As for real
galaxies, the angular momentum of the halos tend also to be aligned perpendicular to the
radial direction of the void.

Although the orientation of the galaxies and halos within the large-scale structure has been
identified both observationally and numerically, there has not been any previous quantitative
study neither of the relationship between the various kind of alignments already determined,
nor of the connection between these alignments and the existing results for the positioning
of the angular momentum with respect to the halo axes. In this paper we use and discuss
several previous results concerning the alignments of halo axes and angular momentum with
respect to their corresponding voids for those halos lying in the shells of the voids. Not
all these results are independent, so we intend to establish a convenient set of independent
results from which all other results may be obtained. To this end we shall present the needed
mathematical identities relating the various kind of alignments that we shall consider, and
use them to infer the set of independent results that we shall choose. Then we shall show
that, using the appropriate mathematical relationships, all previous results can be derived,
thereby proving their consistency. In this way we shall not only present previous results in
a compact and ordered fashion, but shall also present results whose form is valid for the
alignment with respect to features of the large scale structure other than voids.

In Section 2, we show how to characterise uniquely the statistics of the orientation of the
halos in the shells of the voids with respect to their corresponding voids and how the
alignments of the various halo axes may be derived from it. In Section 3, we provide a
motivated model for the probability distribution for the orientation of the angular momentum
of a randomly chosen halo with respect to its axes. Then, through a modification of this
model, we obtain a model for the alignment of the angular momentum with respect to the halo
axis for a halo in the neighbourhood of a void (or, in general, within a large scale tidal
field). Using this model along with the statistics for the orientation of the halo with
respect to the void we obtain the probability distribution for the alignment of the angular
momentum with the radial direction of the void. Then, comparing this distribution with that
found in simulations, we calibrate that model, inferring a very substantial dependence of the
alignment of the angular momentum with respect to the halo axes on the environment. Finally,
in Section 4, we discuss  the relationship between the alignment of the angular momentum of
the halo at the moment of turn around and at present, and how any uncertainty in the
determination of the angle of the rotation axis of galaxies with respect to the radial vector
joining the centre of the corresponding void with the galaxy diminishes the strength of the
alignment. A discussion and a summary of conclusions are given in Sec. 5.

\section{Alignments of halo axes with respect to the void radial direction}

We shall first consider the alignments of the three halo axes with respect to the void radial
direction. These alignments were quantified first by Patiri et al. (2006) and, with better
statistics, by Brunino et al. (2007) and Cuesta et al. (2008). In this last work it was
argued that all positional information of halos with respect to the void may be encoded by
the three Euler angles ($\theta$, $\psi$, $\phi$) that transform the three axes of the halo
into the local orthogonal base (i.e. the void radial direction, and the directions along
parallels and meridians for any arbitrarily chosen set of geographical coordinates on the
shell of the voids).

We call $\theta$ the angle between the major axis and the radial vector joining the centre of
the void with the centre of the  dark matter halo, $\psi$ the angle between the direction
perpendicular to the major axis within the plane of the void and the minor axis, and $\phi$
the angle between the direction perpendicular to the major axis within the plane of the void
and the direction along the local parallel (see curve A in Fig. 1). Given the isotropy of the
problem within the plane of the void, $\phi$ must be uniformly distributed between 0 and
$\pi$. This was checked in simulations, for consistency, by Cuesta et al. (2008). Note that
since these Euler angles determine the position of three orthogonal axes rather than a set of
three orthogonal vectors, we have $\theta$$\in$[0,$\pi$/2], $\psi$$\in$[0,$\pi$] and
$\phi$$\in$[0,$\pi$].

\begin{figure}
\epsfig{file=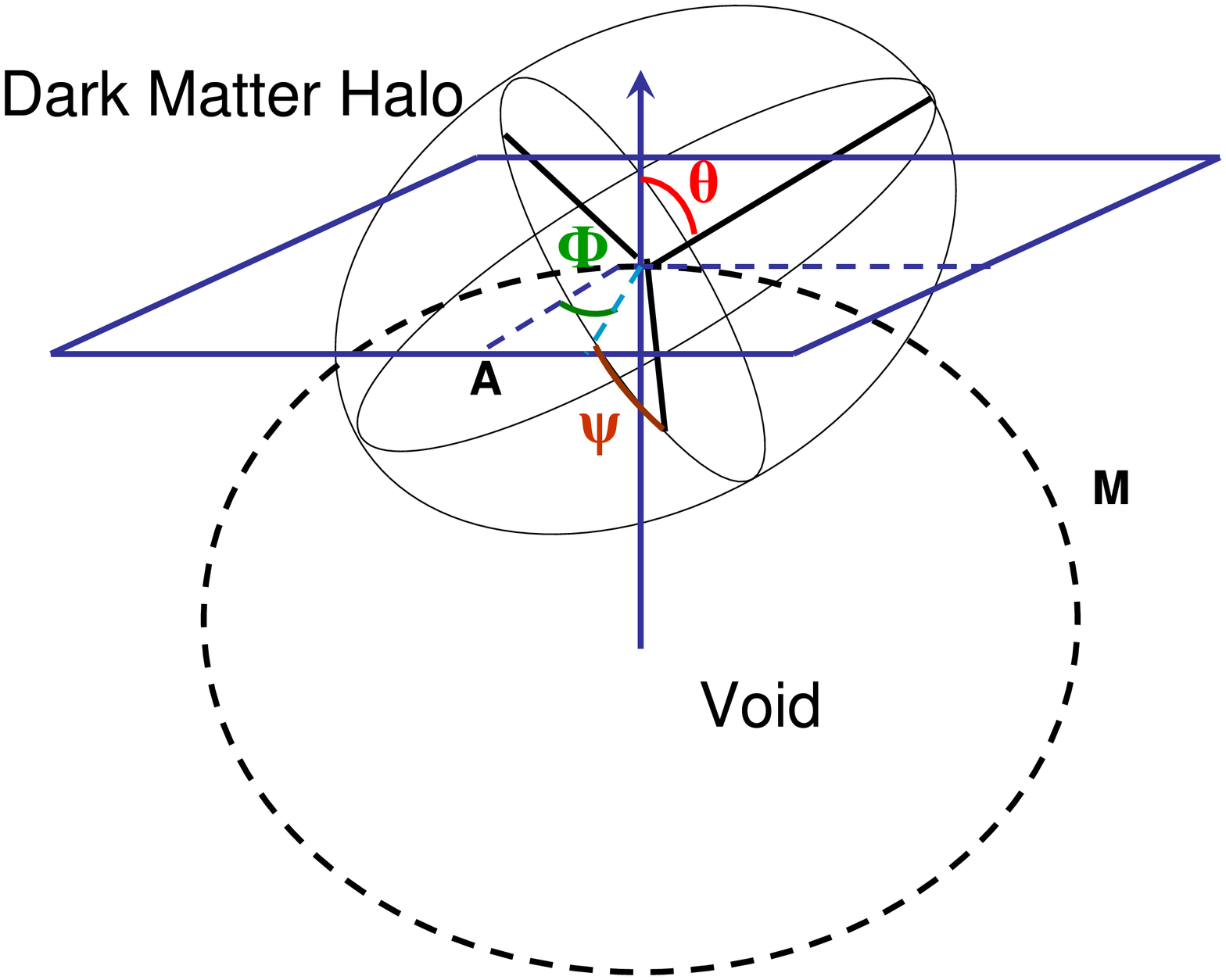,width=0.5\textwidth, angle=-90}

 \caption{Schematic illustration of the orientation of the dark matter halo in the void
 surface. The position of three axes of the dark matter halo are characterised by the
 Euler angles ($\theta$, $\psi$, $\phi$). A, M are the local parallel and meridian,
 respectively.}
\end{figure}

If the shape of the halos were not aligned with any particular direction, $\cos\theta$,
$\psi$, and $\phi$ will be distributed uniformly. However, the numerical simulations have
shown that halos are not oriented randomly within the large scale structure. The
probability distribution of $\theta$ has been explored by mean of several simulations. In
the particular case of halos within a shell with a thickness of 4 h$^{-1}$Mpc around voids
larger than  10 h$^{-1}$Mpc (defined by not containing halos with masses larger than 7.6
$\times$10$^{11}$h$^{-1}$M$_{\sun}$), the simulations have shown that this probability
distribution can be well described by the following expression:

\begin{equation}
P(\mu)d\mu=\frac{p d\mu}{[1+(p^2-1)\mu^2]^{3/2}}; \mu\equiv cos\theta
\label{probcos}
\end{equation}

with p a free parameter that is found to be 1.111$\pm$0.004 (see e.g. Cuesta et al.
2008). Cuesta et al. (2008) also have studied in the simulations the distribution of
the angle $\psi$. They found that the distribution have a small, but highly significant,
departure from uniformity. Given the small amplitude of this non-uniformity, we propose
the following expression containing the simplest $\psi$ dependent term which is even
with respect to $\psi=\pi$/2. This is a condition that must be satisfied due to the
symmetry of the problem.

\begin{equation}
P(\psi)d\psi=\frac{2}{\pi}(1+\beta\cos(2\psi))d\psi
\label{probpsi}
\end{equation}

This expression is able to describe fairly well the data from the simulations with
$\beta$=-0.093$\pm$0.005. This means that it is more likely to find
the minor axis perpendicular to the intersection of the plane of the void with the
plane defined by the middle and minor axes than parallel to this intersection. The
opposite is obviously true for the middle axis.

Note that in the above expression $\psi$ is assumed to take values between 0 and
$\pi$/2. As we mentioned above, $\psi$ actually goes between 0 and $\pi$, but due to
the symmetry of P($\psi$) around $\psi$=$\pi$/2 it is not necessary to consider values of
$\psi$ larger than $\pi$/2 (even though the position $\psi$ is not, in general, equivalent to
$\pi$-$\psi$). However, we would like to stress that in the process of deriving the
forthcoming mathematical identities $\psi$ is assumed to go from 0 to $\pi$ and it is
 only in the final expressions that the symmetry of P($\psi$) is exploited, so that
in all the final expressions $\psi$ is assumed to be between 0 and $\pi$/2.

From the two expressions we have stated before, it follows that the full statistical
information about the orientation of the halos with respect to the void is encoded in the
joint probability distribution for $\cos(\theta)$ and $\psi$:
P$_{or}$($\cos\theta$,$\psi$). We drop the dependence on $\phi$ since it is trivial.  We
propose the following expression to combine the above two probabilities, P($\cos\theta$)
and P($\psi$):

\begin{eqnarray}
P_{or}(\cos\theta,\psi)d\cos\theta
d\psi=\left(\frac{p}{[1+(p^2-1)\cos^2\theta]^{3/2}}\right)
\nonumber\\
\left(\frac{2}{\pi}(1-\beta'\sin^2\theta+2\beta'\sin^2\theta\cos^2\psi)\right)
d\cos\theta d\psi ; \theta, \psi \in [0,\pi/2] 
\label{probcospsi}
\end{eqnarray}

The first parenthesis corresponds to the probability distribution for $\cos\theta$ (Eq.
\ref{probcos}) and the second parenthesis to the conditional probability distribution
P($\psi$$\mid$$\theta$) for $\psi$  for a given value of $\theta$. To relate the parameter
$\beta'$ with $\beta$ we applied the fact that, for the halos under consideration (p=1.11),
marginalising the above probability P$_{or}$ with respect to $\cos\theta$,  Eq. \ref{probpsi}
must be obtained. On doing this we obtain: $\beta'$=1.483$\beta$.

We may now note that the same physical cause (the shear of the velocity field)
originates the departure of the distribution of  $\cos\theta$ and $\psi$ from
uniformity. Both effects are of the same order on the shear. Consequently, to leading
order the quantity determining the departure from uniformity of $\cos\theta$ (i.e.
$p-1$) must be proportional to the quantity determining the departure of $\psi$ from
uniformity (i.e. $\beta'$). We may use then Eq. \ref{probcospsi} with:

\begin{equation}
\beta'=-1.22(p-1)
\label{betaexpression}
\end{equation}

From general considerations about the mechanisms generating the alignments one should expect
this expression to be valid not only for the voids used to calibrate this expression but also for
different voids, and also for halo alignments with respect to other large scale
structures.

Using Eq. \ref{probcospsi} we may obtain any statistic concerning the orientation of the
halos with respect to the voids. For example, the alignment of the minor and middle axes
with the radial direction (that has been already measured in simulations) can be expressed
in terms of expression \ref{probcospsi}. The probability distribution,
P$_{min}$($\cos\gamma$), for the cosine of the angle $\gamma$ between the minor axis and
the radial direction of the void is related to P$_{or}$ by the following expression:

\begin{equation}
P_{min}(\cos\gamma)=\int_{0}^{1}\int_{0}^{1}P^{'}_{or}(\cos\theta,\cos\psi)
\delta(\sin\theta\sin\psi-\cos\gamma)d\cos\theta d\cos\psi
\label{pminprob}
\end{equation}

where the Dirac's $\delta$ distribution implement the constraint (based on spherical
geometry) that only those values of $\theta$ and $\psi$ corresponding to an orientation of
the halo in which the cosine of the angle between the minor axis and the radial direction
take the value $\cos\gamma$ contribute to the integral. P$^{'}_{or}$ is simply the
distribution implied by P$_{or}$ for the variables $\theta$ and $\psi$:

\begin{equation}
P_{or}(x,y)\equiv\frac{P^{'}_{or}(x,\arccos y)}{\sqrt{1-y^2}}
\end{equation}

Carrying out the integral over $\psi$ we find:

\begin{equation}
P_{min}(\cos\gamma)=\cos\gamma\int_{\pi/2-\gamma}^{\pi/2}P^{'}_{or}
(\cos\theta,\frac{\sqrt{\sin^2\gamma-\cos^2\theta}}{\sin\theta})
\frac{d\theta}{\sqrt{\sin^2\gamma-\cos^2\theta}}
\end{equation}

In a similar fashion, simply changing $\sin\psi$ by $\cos\psi$ in the
argument of the Dirac $\delta$ distribution, we find for the probability distribution,
P$_{mid}$($\cos\eta$), for the cosine of the angle $\eta$ between the middle
axis and the radial direction:

\begin{equation}
P_{mid}(\cos\eta)=\int_{\pi/2-\eta}^{\pi/2}P^{'}_{or}
(\cos\theta,\cos\eta/\sin\theta)d\theta
\label{pmidprob}
\end{equation}

\begin{figure}

\epsfig{file=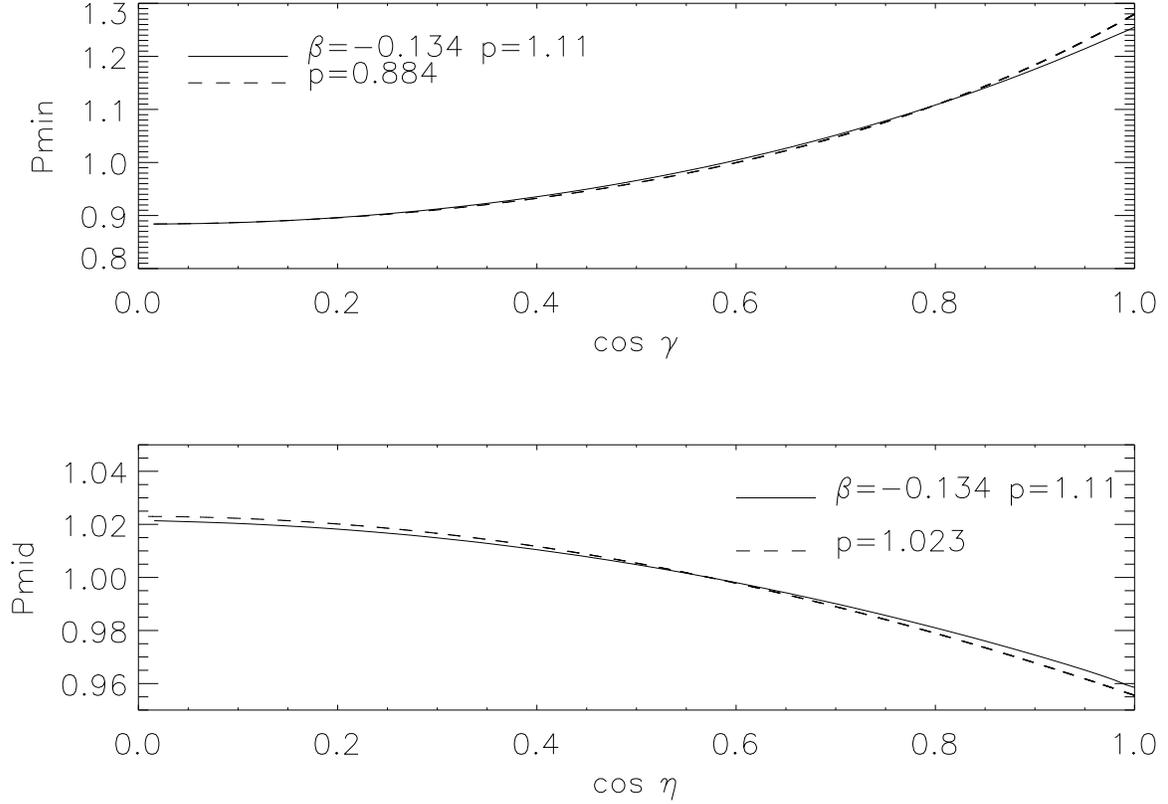,width=\textwidth, angle=0}

\caption{$Upper$ $panel$. The solid line is the probability distribution,
P$_{min}$(cos$\gamma$), for the cosine of the angle $\gamma$ between the minor axis and the
radial direction of the void as given by expression \ref{pminprob} with  $\beta^{'}$=-0.134
and p=1.11 (see text for details). The dashed line is the analytic fit to the results from
the simulations by Cuesta et al. (2008) given by Eq. 1 using p=0.884. $Lower$ $panel$. The
solid line is probability distribution, P$_{mid}$(cos$\gamma$), for the cosine of the angle
$\gamma$ between the middle axis and the radial direction of the void as given by expression
\ref{pmidprob} of $\beta^{'}$=-0.134 and p=1.11 (see text for details). The dashed line is
the analytic fit to the results from the simulations by Cuesta et al. (2008) given by Eq. 1
using p=1.023.}

\label{pminmed}

\end{figure}

Using here expression \ref{probcospsi} with $\beta^{'}$=-0.134 and p=1.11, which is the value
corresponding to the halos studied by Cuesta et al. (2008) we recover (see Fig. \ref{pminmed})
the distributions $P_{mid}$ and $P_{min}$ determined directly in that work.

\section{Alignment of the angular momentum of the halo with respect to the halo axes}

The alignment of the angular momentum of the halo with respect to their axes have been
studied in detail in simulations (Bailin \& Steinmetz 2005; Gottl\"ober \& Turchaninov 2006; Patiri et
al. 2006). We shall now consider a simple analytic model able to fit these alignments.
To this end we take into account a set of halos with principal moments of inertia I$_1$,
I$_2$ and I$_3$. Assuming that halos can interchange rotational kinetic energy and that
they have reached statistical equilibrium, the probability distribution for the three
components of the angular momentum J$_1$, J$_2$ and J$_3$ along the principal axes is
proportional to the Boltzmann factor:

\begin{equation}
P(\vec{J})\propto\exp{\left\{-\left(\sum_{i=1}^3\frac{J_i^2}{2I_i}\right)/kT\right\}}
\end{equation}

Let's write the components of $\vec{J}$ in terms of its modulus J, the angle
$\theta_J$ between  $\vec{J}$ and the halo major axis, and the angle $\phi_J$
between the projection of  $\vec{J}$ onto the plane perpendicular to the halo major
axis and the halo minor axis. Integrating over J we find for the
probability distribution P$_{\bf{J}}$($\cos\theta_J$,$\phi_J$):

\begin{eqnarray}
P_{\bf{J}}(\cos\theta_J,\phi_J)=\frac{3}{2}\frac{2}{\pi\sqrt{\bar{I}_2\bar{I}_3}}
\nonumber\\
\frac{1}{\left[\bar{I}_2^{-1}+\left(\bar{I}_3^{-1}-\bar{I}_2^{-1}\right)\cos^2\phi_J
+\left(1-\bar{I}_2^{-1}-\left(\bar{I}_3^{-1}-\bar{I}_2^{-1}
\right)\cos^2\phi_J\right)\cos^2\theta_J\right]}
\label{probang}
\end{eqnarray}

with $\bar{I}_2$$\equiv\frac{I_2}{I_1}$ and $\bar{I}_3$$\equiv\frac{I_3}{I_1}$
and $\theta_J\in[0,\pi/2]$ and $\phi_J\in[0,\pi/2]$.

If the distribution of rotational kinetic energy amongst halos were in statistical
equilibrium, setting in this expression the mean values of $\bar{I}_2$ and $\bar{I}_3$
($\bar{I}_2$=1.38;  $\bar{I}_3$=1.58, derived from the results of Jing \& Suto 2000) one
would obtain a good approximation to the actual distribution, since the dispersions of
$\bar{I}_2$ and $\bar{I}_3$ are not too large. However, this is not what is observed in the
simulations, because the actual distribution of the angular momentum is far from statistical
equilibrium. Nonetheless, Eq. \ref{probang} can be used as a good model for the angular
momentum distribution found in the simulations using the following values: $\bar{I}_2$=2.74
and  $\bar{I}_3$=7.41. On doing this, we find: $<\cos\theta_{J1}>$=0.315; 
$<\cos\theta_{J2}>$=0.471 and $<\cos\theta_{J3}>$=0.671 where $\theta_{J1}$, $\theta_{J2}$
and $\theta_{J3}$ stands, respectively, for the angle between the angular momentum vector and
the major, middle and minor halo axes. These numbers can be compared with the values found in
the simulations for randomly chosen halos: 0.32, 0.48 and 0.69 (Cuesta et al. 2008). In those
simulations it has been also found that above 75\% of halos have $\theta_{J1}$ larger than
60 degrees. Eq. \ref{probang} predicts 73.5\%.

\subsection{Alignment of the angular momentum of the halo with respect to the
halo axes for a given orientation of the halo with respect to the void}

The above expression gives the probability distribution for the orientation of $\vec{J}$
with respect to the halo axes regardless of the orientation of the halo itself with respect
to the large scale structure. However,  from theoretical considerations we know that the distribution of $\theta_{J}$ and $\phi_{J}$ depends
on the orientation of the halo. We shall show later (in the next subsection) that for halos
in the shell of the voids considered by Cuesta et al. (2008), the probability distribution
P$_{\bf{J}}$($\theta_J$,$\phi_J$$\mid$$\theta$,$\psi$) for the angles $\theta_{J}$ and $\phi_{J}$
in those halos whose major axis form an angle $\theta$ with the radial direction of the
void may be approximated by:

\begin{eqnarray}
P_{\bf{J}}(\cos\theta_J,\phi_J\mid\theta,\psi)=\frac{0.222}{\pi/2}
\nonumber\\
\frac{(F1\times F2\times F3)^{-1/2}}{[0.365F2+(0.135F3-0.365F2)\cos^2\phi_J+
(F1-0.365F2-(0.135F3-0.365F2)\cos^2\phi_J)\cos^2\theta_J]^{3/2}}
\label{probcosphitheta}
\end{eqnarray}

with F1$\equiv$1+$\alpha$($\cos^2\theta$-1/3),
F2$\equiv$1+$\alpha$($\sin^2\theta$$\cos^2\psi$-1/3), 
F3$\equiv$1+$\alpha$($\sin^2\theta$$\sin^2\psi$-1/3) and $\alpha$=-0.68
($\theta_J\in[0,\pi/2]$ and $\phi_J\in[0,\pi/2]$). This expression implies that for
$\theta$=0 and any value of $\psi$ (which corresponds to have the major axis oriented in the
void radial direction with the other two axes lying in the void plane):
$<\cos\theta_{J1}>$=0.251;  $<\cos\theta_{J2}>$=0.489 and $<\cos\theta_{J3}>$=0.693. For
$\theta$=$\pi$/2 and $\psi$=$\pi$/2 (which corresponds to have the major and the middle axes
in the plane of the void and the minor axis oriented in the void radial direction)  we find
$<\cos\theta_{J1}>$=0.354;  $<\cos\theta_{J2}>$=0.526 and $<\cos\theta_{J3}>$=0.598.

From the above numbers we see that the mean values of $<\cos\theta_{J1}>$, being
$\theta_{J1}$ the angle between $\vec{J}$ and the halo major axis, depends very substantially
on the orientation of the halo with respect to the void. To understand the origin of this
dependence we only need to consider the following: the axis corresponding to the smallest
proper value of the shear of the velocity field associated with the void expansion is
parallel to the radial direction of the void. This produces an extra component of $\vec{J}$
which is perpendicular to that direction. Consequently, when  the halo major axis is parallel
to the radial direction ($\theta$=0), the extra component of $\vec{J}$ is also
perpendicular to the halo major axis. So, the average value $<\cos\theta_{J1}>$ for this
particular configuration  must be smaller than the average value over all halos:  
$<\cos\theta_{J1}>$=0.315. On the other hand, when $\theta$=$\pi$/2, the tendency of 
$\vec{J}$ to lay parallel to the surface of the void does not imply an obvious tendency for
$<\cos\theta_{J1}>$. But, to compensate for the smaller value of $<\cos\theta_{J1}>$ when
$\theta$=0, the mean value of $<\cos\theta_{J1}>$ when $\theta$=$\pi$/2 must be larger
than for a randomly chosen halo.

The value of $\alpha$ given in expression \ref{probcosphitheta} corresponds to the halos
considered by Cuesta et al. (2008), but that expression can be made universally valid by
simply expressing $\alpha$ as a function of p (the parameter measuring the strength of the
alignment of the major axis with the radial direction):

\begin{equation}
\alpha=-8.64(p-1)
\label{alphaexpression}
\end{equation}

This relation should be valid in a general situation (with some reservations as presented in
the discussion section) where both the orientation of the
halos and their angular momentum are affected by a large scale shear of the velocity field.

It must be noted that, unlike Eq. \ref{probang}, expression \ref{probcosphitheta} has not
been checked with numerical simulations. It has been derived from Eq. \ref{probang} assuming
a simple model for the dependencies on $\theta$ and $\psi$ and fitting the value of $\alpha$.
In this respect it may be considered a prediction for the dependence on $\theta$ and $\psi$
of the probability distribution for $\theta_J$ and $\phi_J$.

\subsection{Alignment of the angular momentum of the halo with respect to the
void radial direction}

Having determined the probability distribution for the orientation of the halo with respect
to the void, $P_{or}(\cos\theta,\psi)$, and the probability distribution for the orientation
of the angular momentum with respect to the halo for a given orientation of the halo with
respect to the void, $P_{\bf{J}}(\cos\theta_J,\phi_J\mid\theta,\psi)$, we can now express the
probability distribution  $P_{\bf{J}}(\cos\mu)$ for the angle between $\vec{J}$ and the
radial direction of the void in the following manner:

\begin{eqnarray}
\bar{P}_{\bf{J}}(\cos\mu)=\frac{2}{64}\int_{-1}^{1}\int_{-1}^{1}\int_{0}^{2\pi}\int_{0}^{2\pi}
P_{or}(\cos\theta,\psi)P_{\bf{J}}(\cos\theta_J,\phi_J\mid\theta,\psi)
\nonumber\\
\delta(\cos\theta\cos\theta_J+\sin\theta\sin\theta_J\cos(\psi-\phi_J)-\cos\mu)
d\phi_Jd\psi d\cos\theta d\cos\theta_J
\label{pcosmu}
\end{eqnarray}

The factor 1/64 comes from the fact that the distributions $P_{or}$ and $\bar{P}_{\bf{J}}$
are normalised with all angles between 0 and $\pi$/2. This is because, as long as we consider
only the orientation of halos or the orientation of  $\vec{J}$ within the halo, given the
symmetries, this is the meaningful range for the angles. However, when we consider both
orientations simultaneously to obtain the position of $\vec{J}$ with respect to the void we
must consider all angles between 0 and $\pi$ to cover all possible configurations. The factor
2 comes from normalising $P_J(\cos\mu)$ to the range $\cos\mu\in$[0,1].

Integrating over $\phi_J$ and taking advantage of the symmetries of the integral, we find:

\begin{eqnarray}
\bar{P}_{\bf{J}}(\cos\mu)=\frac{1}{8}
\int_{0}^{\pi}\int_{-1}^{1}\int_{\cos(\theta+\mu)}^{\cos(\theta-\mu)}
\nonumber\\
\frac{P_{or}(\cos\theta,\psi)P_{\bf{J}}(\cos\theta_J,\phi_J\mid\theta,\psi)
d\cos\theta_Jd\cos\theta d\psi}
{\sqrt{(\sin\theta\sin\theta_J)^2-(\cos\mu-\cos\theta\cos\theta_J)^2}}
\label{pjcosmu}
\end{eqnarray}

with $\phi_J\equiv\arccos((\cos\mu-\cos\theta\cos\theta_J)/(\sin\theta\sin\theta_J))+\psi$.

\begin{figure}

\epsfig{file=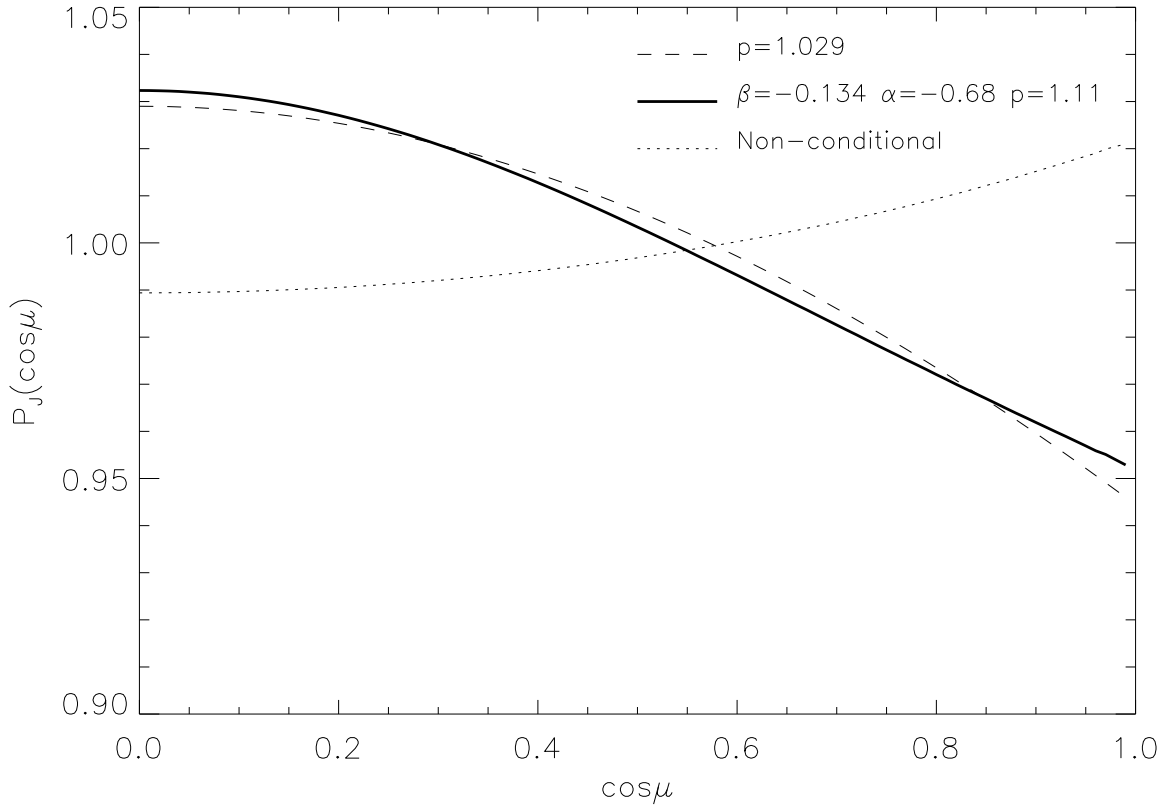,width=\textwidth, angle=0}

\caption{The thick solid line is the probability distribution, $\bar{P}_{\bf{J}}$($\cos\mu$),
for the cosine of the angle $\mu$ between the angular momentum vector of the halo $\vec{J}$
and the radial direction of the void as given by Eq. \ref{pjcosmu} with P$_{or}$ given by Eq.
\ref{probcospsi} and P$_J$ given by Eq. \ref{probcosphitheta}; The dot line represents the
unconditional model which is estimated as above but using all F's equal to one in
Eq.\ref{probcosphitheta}. The dashed line is the analytic fit to the results from the
simulations by Cuesta et al. (2008) given by Eq. 1 using p=1.029.}

\label{pjcosmufig}

\end{figure}

If we assumed that the orientation of the angular momentum with respect to the halo is not
affected by the orientation of the halo with respect to the void, the conditional
distribution,  $P_{\bf{J}}(\cos\theta_J,\phi_J\mid\theta,\psi)$, would be equal to the
unconditional one $P_{\bf{J}}(\cos\theta_J,\phi_J)$ given by Eq. \ref{probang} with
$\bar{I}_2$=2.74 and  $\bar{I}_3$=7.41. Then using Eq. \ref{probcospsi} with
$\beta^{'}$=-0.134 for $P_{or}$, which corresponds to the halos around the large voids
studied by Cuesta et al. (2008), and inserting both distributions into expression
\ref{pcosmu} we would not recover the distribution for $\cos\mu$ found in the simulations,
which is well parametrised by:

\begin{equation}
P_{\bf{J}}(\cos\mu)=\frac{p}{(1+(p^2-1)\cos^2\mu)^{3/2}}
\label{pobserved}
\end{equation}

with p=1.029. Instead, we obtain an alignment with the opposite sign (see Fig.
\ref{pjcosmufig}).

To reproduce the observed distribution we must allow for a dependence of $P_{\bf{J}}$ on
$\theta$ and $\psi$. A simple model for this dependence has been given in the previous
subsection (see Eq. \ref{probcosphitheta}). Inserting this into expression \ref{pjcosmu}
together with expression \ref{probcospsi} with $\beta^{'}$=-0.134 for $P_{or}$ we may fit
$\alpha$ so as to obtain the best fit (with $\alpha$=-0.68) to the numerical results which
are well described by Eq. \ref{pobserved}. This is how we have obtained the value of $\alpha$
given in \ref{probcosphitheta}. It must be noted that this expression represent a model in
which, for any value of $\theta$, $\psi$ the alignment of $\vec{J}$ with the axes follows
expression \ref{probang} (corresponding to randomly chosen halos), but where the parameters
show a small dependence on $\theta$ and $\psi$ (the simplest with the right symmetry).

 The fact that expression \ref{pjcosmu}, with expression \ref{probcospsi} and
\ref{probcosphitheta} for P$_{or}$, P$_{\bf{J}}$, respectively, agrees with \ref{pobserved}
to a high degree of accuracy for any value of $\cos\mu$ confirms the validity of this model
for the conditional $P_{\bf{J}}$ (see Fig.
\ref{pjcosmufig}).

If the two distributions entering expression \ref{pjcosmu} were independent of $\psi$ and
$\phi_J$, i.e. if they take the following form:

\begin{eqnarray}
P_1(\theta,\psi)=\frac{1}{\pi}P_1(\cos\theta)
\nonumber\\
P_2(\theta,\psi)=\frac{1}{\pi}P_2(\cos\theta_J\mid\theta)
\end{eqnarray}

both normalised with all angles between 0 and $\pi$, then expression \ref{pjcosmu}, after
integrating over $\psi$, reduces to:

\begin{equation}
P(\cos\mu)=\frac{1}{2\pi}\int_{-1}^1\int_{\cos(\theta+\mu)}^{\cos(\theta-\mu)}
\frac{P_1(\cos\theta)P_2(\cos\theta_J\mid\theta)  d\cos\theta_J d\cos\theta}
{\sqrt{\sin^2\theta\sin^2\theta_J-(\cos\mu-\cos\theta\cos\theta_J)^2}}
\label{pconvol}
\end{equation}

This expression gives the probability distribution for the angle $\mu$ between a fixed vector
$\vec{H}$ and a vector $\vec{a}$ determined as follows: choose a vector $\vec{b}$ which forms
and angle $\theta$ with $\vec{H}$ with a probability distribution given by P$_1$ and a
randomly chosen (uniformly) azimuthal angle. Then, choose the vector $\vec{a}$ forming an
angle $\theta_J$ with $\vec{b}$ with a probability distribution given by P$_2$ and a randomly
(uniformly) distributed azimuthal angle.

In the case where P$_2$ does not depend on $\theta$ this expression is symmetric in P$_1$
and P$_2$. This is not evident at first sight, given the different integration limits
for $\theta$ and $\theta_J$. $P(\cos\mu)$ can be seen as the two-dimensional convolution on
the sphere of P$_1$ and P$_2$. We will use this expression in the next section to study
the effect of the dilution of the signal of the alignments due to: a) the miss-aligning
processes after the moment of turn-around and b) the  observational errors in the
estimation of $\theta$.

\section{Degradation of the alignments}

\subsection{Degradation of the alignments due to the miss-aligning process after the
moment of turn-around}

In the standard (linear) tidal torque theory (TTT), the torque acting on a protohalo keeps
its direction invariant through the time. It has been found in numerical simulations that
the TTT predicts correctly the angular momentum of the protohalo up to the moment of the
turn-around (Porciani et al. 2002). Hereafter, the tidal torque starts to rotate so 
by the time the halo virialises, its angular momentum will form an angle $\zeta$ with respect to
the initial torque. The initial torque direction is close to the direction of the angular
momentum at the moment of the turn-around. This evolutionary process has been studied in
detail by Porciani et al. (2002). We have found that the
distribution of the angle $\zeta$ that they found can be approximated by:

\begin{equation}
P(\zeta)\simeq16.83[e^{-\sin\zeta/0.21}(1-\Theta(\zeta-\pi/2))+0.0085\Theta(\zeta-\pi/2)]\sin\zeta
\label{pzeta}
\end{equation}

where $\Theta$ is the Heaviside function.

Assume now that we have derived the theoretical predictions for the alignment of $\vec{J}$
with the large scale structure under the assumption of a non-rotating torque, whose
probability distribution we represent by $P^{'}_{\bf{J}}$ (that turns out to be given by Eq.
\ref{pobserved} with a certain value of p). To obtain the probability distribution,
$P_{\bf{J}}$, corresponding to the alignment of the actual $\vec{J}$ we must  take into
account the fact that the angular momentum of the halo is rotated by an angle $\zeta$ with
respect to the theoretical prediction. Consequently,P$_{\bf{J}}$ may be expressed as
the convolution on the sphere of $P^{'}_{\bf{J}}$ and 2$P(\zeta)$, which is given by
expression \ref{pconvol} with these two distributions in the place of P$_1$ and P$_2$. The
factor 2, in 2$P(\zeta)$, is due to the fact that $P(\zeta)$, which is not symmetric with
respect to $\pi$/2, is not normalised within the interval (0,$\pi$/2), unlike the
distribution appearing in Eq. \ref{pconvol}.

\begin{figure}

\epsfig{file=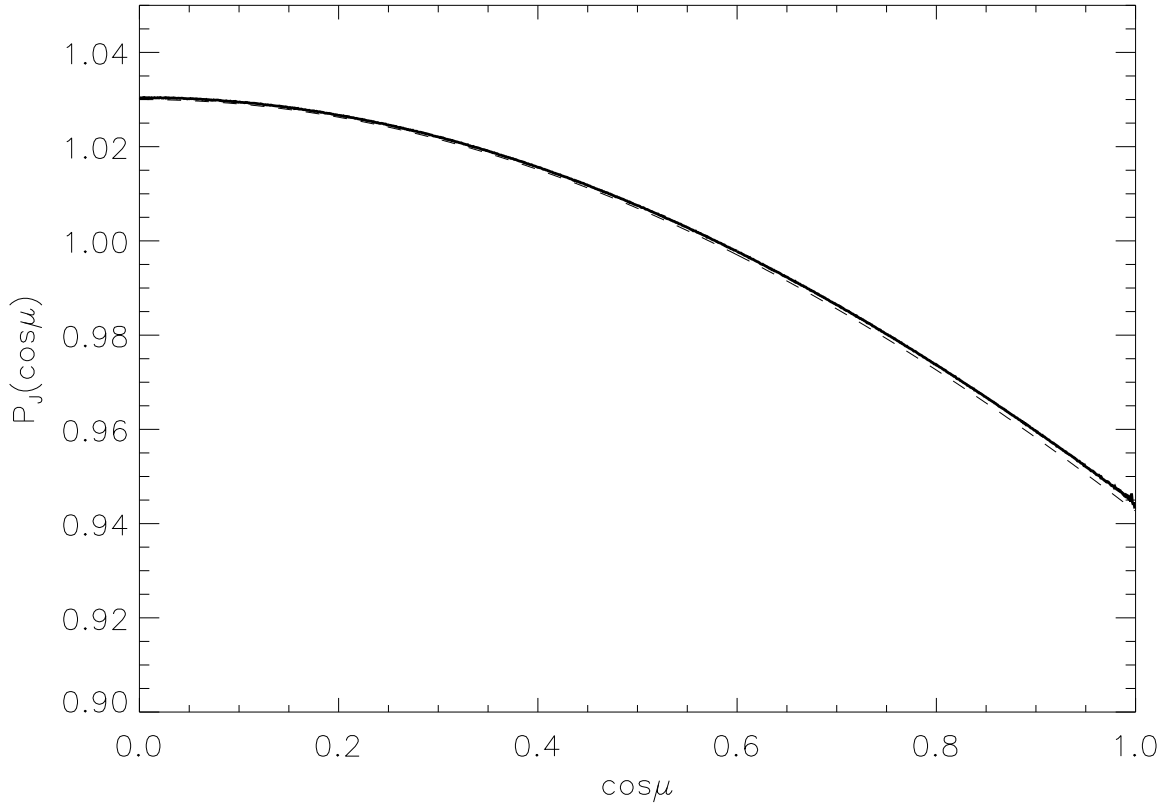,width=\textwidth, angle=0}

\caption{The thick solid line is the probability distribution, ${P}_{\bf{J}}$($\cos\mu$), for
the cosine of the angle $\mu$ between the angular momentum vector of the halo $\vec{J}$ and
the radial direction of the void  accounting for the tidal torque rotation during and after
the formation of the halo. To estimate the effect of the rotation of the tidal torque, we
have  assumed that, under the assumption of a non-rotating torque, ${P}_{\bf{J}}$($\cos\mu$)
is given by Eq. \ref{pobserved} with p=1.06 and we have convolved this expression with Eq.
\ref{pzeta} using Eq. \ref{pconvol}. The dashed line is the analytic fit to the results from
the simulations by Cuesta et al. (2008) given by Eq. 1 using p=1.03. We find that the
strength of the alignment, which is measured by p-1, reduces to a half with respect to the
theoretical prediction under the assumption of a non-rotating torque.}

\label{spinrotation}

\end{figure}

If we use for $P^{'}_{\bf{J}}$ a value of p=1.06, the resulting $P_{\bf{J}}$ (see Fig.
\ref{spinrotation})
is also well approximated by expression \ref{pobserved} with a value of p of 1.03. So, we
see that the strength of the alignment, which is measured by p-1, reduces to a half with
respect to the theoretical predictions with a non rotating torque. This ratio between the
value of p-1 for  $P^{'}_{\bf{J}}$ and $P_{\bf{J}}$ is constant to a very good approximation
for all interesting p values (p-1$\ll$1).

Note that in obtaining $P_{\bf{J}}$ from $P^{'}_{\bf{J}}$ and $P(\zeta)$ we have assumed
that the rotation of $\vec{J}$ with respect to its primordial direction is independent of
the orientation with respect to the large scale structure. This is a good approximation,
because the large scale makes up only a small contribution to the total tidal field, and
this contribution rotates less than that due to smaller scales.

\subsection{Degradation of the alignment due to observational errors at estimating $\theta$}

When measuring alignments between the large scale structure and the rotational axis of
spiral galaxies, the error in the estimation of the alignment angle due to uncertainties in
the position of the galaxies and in the orientation of its axis leads also to a degradation
of the intrinsic alignment.

To quantify the relevance of this potential source of degradation of the alignment, we
assume that the probability distribution, $P_{er}(\zeta)$, for the angle $\zeta$ between the
observed and the actual direction of $\vec{J}$ takes the form:

\begin{equation}
P_{er}(\zeta)=\frac{e^{\frac{-\zeta^2}{2\sigma^2}}\sin\zeta}
{\int_0^\pi e^{\frac{-\mu^2}{2\sigma^2}}\sin\mu d\mu}
\label{errorfun}
\end{equation}

\begin{figure}

\epsfig{file=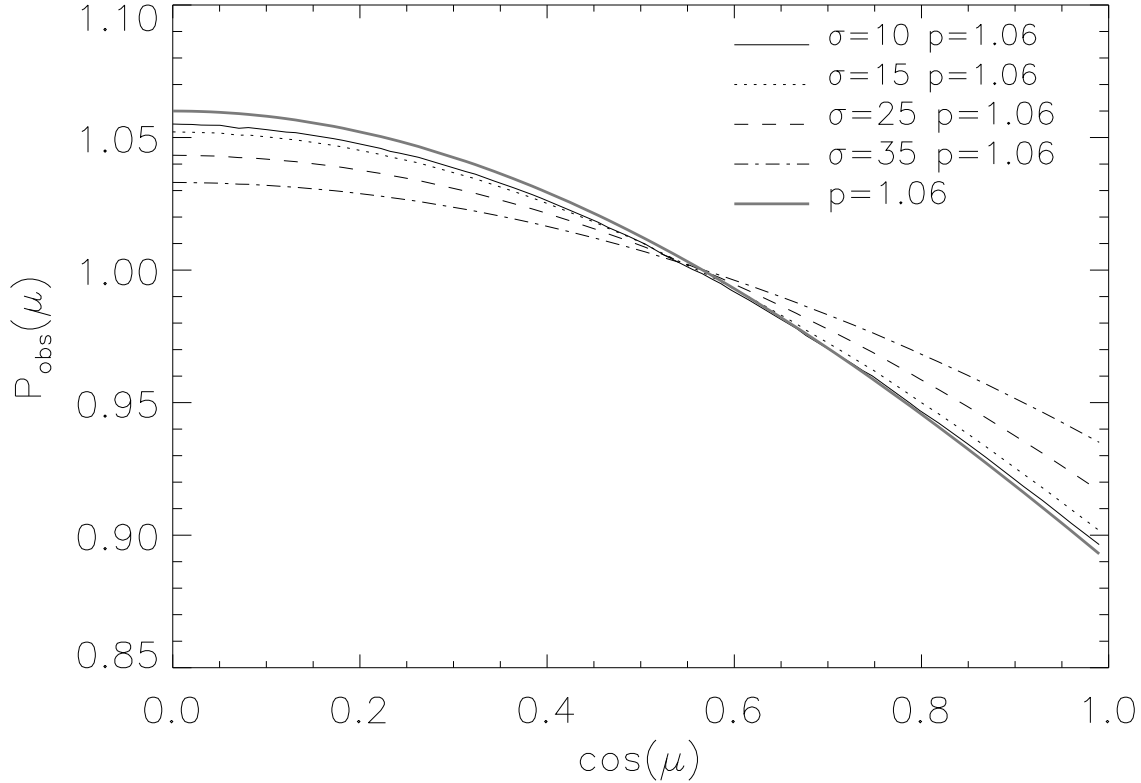,width=\textwidth, angle=0}

\caption{The different lines shows the probability distribution, $P_{obs}(\mu)$, for the
angle $\mu$ between the observed angular momentum vector of the spiral galaxies $\vec{J}$ and
the radial direction of the void assuming four different levels of uncertainty on the angular
position of $\vec{J}$ ($\sigma$=10; $\sigma$=15; $\sigma$=25 and $\sigma$=35 degrees). The
distribution for the intrinsic alignment of the angular momentum is given by Eq.
\ref{pobserved} with p=1.06, whereas the observed distribution is characterised by the
convolution, as provided by Eq. \ref{pconvol}, between the intrinsic signal and the error
function given by Eq. \ref{errorfun}.}

\label{spindegradation}

\end{figure}

Assuming again that the distribution of the angle $\zeta$ is independent of the large scale
structure, we may insert in expression \ref{pconvol}, 2$P_{er}(\zeta)$ and the probability
distribution for the intrinsic alignment of the axes, $P_{in}(\cos\mu)$, to obtain the
distribution for the observed $\mu$, $P_{ob}(\cos\mu)$. Using for $P_{in}$ expression
\ref{pobserved} with p=1.06, we have obtained $P_{ob}$ for four values of $\sigma$ (see
Fig. \ref{spindegradation}). Expected typical uncertainties in the observations,
$\sigma\sim$20 degrees, do not produce a strong decrease in the signal.

\section{Discussion and Conclusions}

We have presented a set of general tools for the description of the alignments of halos and
galaxies with respect to the large scale structure where they are embedded. Using these tools
we have done the following:

\begin{itemize}

\item We have provided an accurate model for the conditional probability distribution
function for the orientation of the angular momentum with respect to the halo axis for halos
with a given orientation with respect to the large scale structure. We have fitted the single
parameter of our model so as to obtain, using the appropriate mathematical identity, the
alignment between the angular momentum of the halos in the shells of cosmic voids that has
been found in  simulations. The conditional PDF obtained in this manner shows a strong
dependence on the orientation of the halo with respect to the void. This dependence is
produced by the tidal field on the shell of the voids which causes the minor axis to be more
aligned with the radial direction than in the isotropic case, while the opposite is true for
angular momentum. If that dependence did not exist, that is, if the PDF for the
orientation of the angular momentum with respect to halo axis for halos in the shells of
voids were the same as for randomly chosen halos (i.e. following an unconditional PDF) then
the angular momentum would show the same kind of alignment with the radial direction than the
minor axis and opposite to that found in simulations. In the past, and under the assumption
that the angular momentum PDF is independent of the environment, there has been some
confusion when interpreting the alignments of halo axis and angular momentum with the void
radial direction.

\item We have shown that the strength of the predicted alignment of the angular momentum with
the void radial direction, as given by the TTT, it is twice as large as the one actually
measured in simulations (as parametrised in this work by p-1). To model this degradation of
the signal we have convolved the angular momentum PDF characterised by a certain parameter p
(Eq. \ref{probcos}) with an analytical fit (based on previous results) that describes the PDF
for the angle between the initial torque and the present angular momentum (Porciani et al.
2002). As long as p-1$\ll$1 the resulting distribution is also well described by the
expression given in Eq. \ref{probcos} with a value of p$'$ such that 1-p$'$=(1-p)/2.
Understanding this degradation is a necessary step towards a fully theoretical explanation of
the alignment of the angular momentum with the void radial direction found in simulations. In
a future work (Betancort-Rijo \& Trujillo 2010) we shall show that the TTT predicts, for the
voids under consideration, a value of the strength of p$\simeq$1.06 for the initial
alignment, so a value of p$\simeq$1.03 should correspond to the alignment of the present
angular momentum, as found in simulations.

\item We have also convolved the initial angular momentum PDF with a Gaussian PDF to model
the observational errors on measuring the angle between the angular momentum and the void
radial direction. For real galaxies this error arises both from the uncertainty on estimating
the direction of the galaxy rotation axis as well as from the error on determining the void
radial direction due to the redshift distortion. We have shown that up to a r.m.s. of roughly
15\% the intrinsic alignment is basically unaffected.

\end{itemize}

In the present work we have dealt explicitly with large cosmic voids, but all expressions may
be used whenever the orientation of the halos or galaxies with respect to the large scale
structure (or more precisely, the orientation with respect to the proper axes of the
deformation tensor or the tidal tensor on large scale) is considered. In order to use the
expressions proposed in this paper in more general cases, two considerations are in order.
First, the PDF for the orientation of halos (Eq. \ref{probcospsi}) may show in general a
dependence on $\phi$. Therefore, in any expression involving integrals of P$_{or}$, the
dependence on $\phi$ must be considered and that expression must also be integrated over
$\phi$ and divided by 2$\pi$. Second, the ratios $\beta$$'$/(p-1) (see expression
\ref{betaexpression}) and $\alpha$/(p-1) (see expression \ref{alphaexpression}) have been
determined using halos within certain shells of cosmic voids. However, from general
considerations about the mechanisms generating the alignments it follows that these ratios
should be essentially independent of the voids used (as long as p-1$\ll$1). In other cases
(i.e. walls or filaments) the same considerations lead to the constancy of the first ratio.
The effect described by the second ratio (i.e. the dependence on environment of the
orientation of the angular momentum with respect to the halo axes) does not depend just on the
present deformation tensor, but also on its deformation history (we will show that on
Betancort-Rijo \& Trujillo 2010). Therefore, its value could depend on the kind of structure
considered. This dependence, though, should be expected to be small, but as much as it can
be measured it might provide an interesting test for the standard (and non-standards) model
of the large-scale structure formation. 



\bsp

\label{lastpage}

\end{document}